\def\etal{{\rm et\ al. }}
\def\mpc{{h^{-1}\rm Mpc}}

\def\kms{{\rm km s^{-1}}}
\newcommand\aap{{\em A}\&{\em A}}
\newcommand\aaps{{\em A}\&{\em AS}}
\newcommand\aj{{\em AJ}}
\newcommand\apj{{\em ApJ}}
\newcommand\apjl{{\em ApJ}}
\newcommand\apjs{{\em ApJS}}
\newcommand\araa{{\em ARA}\&{\em A}}

\newcommand\mnras{{\em MNRAS}}

\documentclass[structabstract]{aa}  
\usepackage{graphicx}
\usepackage{txfonts}
\usepackage[sort]{natbib}
\bibpunct{(}{)}{;}{a}{}{,}
\begin{document}

\titlerunning{Cluster of galaxies behind the Galactic Bulge}
\authorrunning{Coldwell et al.}

\title{Confirmation of a cluster of galaxies hidden behind the 
Galactic bulge using the VVV Survey}
\subtitle{}

\author{Georgina Coldwell\inst{1,2}, Sol Alonso\inst{1,2}, Fernanda Duplancic\inst{1,2},
        Maren Hempel\inst{3},Valentin D. Ivanov\inst{4},
        \and
        Dante Minniti\inst{3,5}
        }

\institute{Consejo Nacional de Investigaciones
Cient\'ificas y T\'ecnicas (CONICET), Argentina
           \and
           Departamento de Geof\'isica y Astronom\'ia- Facultad de Ciencias Exactas, F\'isicas y Naturales- Universidad Nacional de San Juan, San Juan, Argentina
           \and
           Instituto de Astrof\'isica, Pontificia Universidad Cat\'olica de Chile, Casilla 306, Santiago 22, Chile
           \and  European Southern Observatory, Alonso de C\'ordova 3107, Vitacura, Casilla 19001, Santiago, Chile  
           \and     Vatican Observatory, Vatican City State V-00120, Italy 
           }
             
\date{Received xxx; accepted xxx}

\abstract
{$Suzaku$ and $Chandra$ X-ray observations detected a new cluster 
of galaxies, Suzaku J1759$-$3450, at a redshift z $=0.13$. It is 
located behind the Milky Way, and the high Galactic dust 
extinction renders it nearly invisible at optical wavelengths.}
{We attempt here to confirm the galaxy cluster with near-infrared 
imaging observations, and to characterize its central member galaxies.}
{Images from the VVV survey were used to detect candidate member 
galaxies of Suzaku J1759$-$3450 within the central region of the 
cluster, up to 350 $kpc$ from the X-ray peak emission. Color-magnitude and color-color diagrams and 
morphology criteria allowed us to select the galaxies among the 
numerous foreground sources.}
{ Fifteen candidate cluster members were found very close to a modeled red-sequence at the redshift of the cluster. 
Five members are extremely bright, and 
one is possibly a cD galaxy. The asymmetry in the spatial distribution of the galaxies respect to the X-ray peak 
emission is an indicator of that this cluster is still suffering a virialization process. }
{Our investigation of Suzaku J1759$-$3450 demonstrates the 
potential of the VVV Survey to study the hidden population of 
galaxies in the Zone of Avoidance.}


\maketitle
%

\section{Introduction}

Clusters of galaxies are the largest gravitationally bound systems in the Universe. 
Their number density and clustering strongly depend on cosmological parameters
providing potential means to constrain the underlying cosmological model 
\citep{Bahcall03,Gladders07,Rozo09}.
Clusters also harbour a large fraction of all galaxies and provide an outstanding 
environment that promotes their chemical evolution and morphological transitions. 
Consequently, the detection of new galaxy clusters enhances the possibility to
unravel important clues about structure, galaxy formation and evolution.

The thermal X-ray emission from the intracluster medium is a 
reliable tracer of galaxy clusters \citep{Gur72}. 
The detection of galaxy clusters using X-ray emission from hot
gas is a popular technique \citep{Romer01,Pierre06,Finoguenov07}.
\cite{Mori13} used this method 
to identify a rich galaxy cluster, Suzaku J1759$-$3450, behind 
the Milky Way bulge from $Suzaku$ and $Chandra$ observations.
The spatial extent of the X-ray emission with a nearly circular shape and a 
radius of $\sim 4$ arcmin, and a bolometric X-ray luminosity of 
$L_x$($r$$<$$r_{500}$)=4.3$\times$10$^{44}$ erg\,s$^{-1}$
suggests that Suzaku J1759$-$3450 is a new cluster of galaxies, at redshift z $=0.13$.
The peak of the surface brightness is located 
at $(RA,Dec)_{J2000.0}$=$(17^h 59^m 17.^s41,−34\degr50^{'}18.^{''}6)$
or $(l,b)$=$(356.3818\degr,-5.4660\degr)$, and the total mass estimate was
2.2$\times$10$^{14}$$M_{\sun}$ with an estimated core radius of $r_c=1.61$ arcmin. 
A precovery observation is available from the ROSAT Bright Source Catalogue \citep{Voges99}. 

Previous optical, near-infrared (NIR), X-ray and H1 radio surveys 
have detected voids, galaxy clusters and superclusters at low 
Galactic latitude \citep{Woudt04,Kraan00}. Recently, a new NIR ESO 
Public Survey called VISTA Variables in V\'ia L\'actea 
\citep[VVV; ][]{Minniti10,Saito12} became available. Although the 
main scientific goals of VVV are to study the galaxy structure 
\citep{Gonza11,Gonza12}, the hidden stellar clusters 
\citep{Minniti11,Borissova11}, and the nearby stars 
\citep{Beamin13,Ivanov13}, its exquisite depth and angular resolution make it an excellent 
tool to find and study extragalactic objects in the Zone of Avoidance.
For example, \cite{Amores12} identified 204 new galaxy candidates from the VVV 
photometry of 1.636 square degrees near the Galactic plane, 
increasing by more than an order of magnitude the surface density 
of known galaxies behind the Milky Way.

In this work, we report a NIR VVV identification of member 
galaxies in the new Suzaku J1759$-$3450 galaxy cluster. 
Color-magnitude and color-color diagrams of the sources in the 
cluster field were used to select candidate cluster members.
We also identified the red-sequence formed by cluster ellipticals and S0 galaxies.
The presence of such red-sequence in the color-magnitude diagram is a 
clear sign of the existence of structure due to early-type cluster galaxy members 
occupying a well defined region in the color space \citep{Visva77}. 
Therefore, the red-sequence has been used as a galaxy clusters detection method
\citep{GY00} whose slope is related with the cluster redshift \citep{Gla98,LC04,Stott09}.
 
The layout of the paper is as follow: The observational data are described in 
section 2. In section 3 the results of the analysed data are shown. The discussion 
and conclusions are presented in sections 4 and 5, respectively.
Throughout the paper, we use 
the following cosmological parameters: $H_0$=$67.3\,\kms\,h^{-1}$, 
$\Omega_M$=0.315, and $\Omega_{\lambda}$=0.685 \citep{Planck13}.

\section{Observational Data}

\subsection{The VVV Survey}

VVV is a new Galactic NIR ESO public survey with the 4.1-m ESO 
VISTA \citep[Visual and Infrared Survey Telescope for Astronomy; 
][]{Emer04,Emer06,Emer10} telescope, located at Cerro Paranal in 
Chile. It is equipped with a wide field NIR camera VIRCAM 
\citep[VISTA InfraRed CAMera; ][]{Dalton06}. The instrument has 
1.65$\degr$ field of view, and it is sensitive over 
$\lambda$=0.9-2.4\,$\mu$m wavelength range. VIRCAM has 16 Raytheon 
VIRGO 2048$\times$2048 HgCdTe science detectors, with a mean pixel 
scale of 0.34\,arcsec\,px$^{-1}$. Each individual detector covers 
$\sim$694$\times$694\,arcsec$^2$ on the sky and offset exposures are taken  
in order to obtain contiguous coverage. 

The VVV footprint covers a total of $\sim$520 square degrees on 
the sky, nearly equally split between the Milky Way bulge and the 
inner Southern disk. The survey has two components: 
(i) $ZYJHK_S$ bands, and 
(ii) up to 80 multi-epoch $K_S$ observations spread over approximately 7-8 
years, for variability and proper motion studies.
More details can be found in \cite{Minniti10} and \cite{Saito12}.
The observations were processed within the VISTA Data Flow System 
(VDFS) pipeline at the Cambridge Astronomical Survey Unit 
\citep[CASU; ][]{Lewis10}.

For our analysis the $ZY$ bands were omitted due to the larger 
extinction at these bands.
The characteristics of the single epoch images in J, H and $K_S$ filters used in 
this work are summarized in Table 1.
Finally, for the $K_S$ band, we have built a deep image of the field combining 30 different 
epochs. This last was possible since multi-epoch observations were obtained, only for 
the $K_S$ band, for the variability goal of the VVV survey.

\begin{table}
\center
\caption{Observation Parameters}
\begin{tabular}{|c c c c c| }
\hline
Filter & Exp.Time & NDIT & Jitters & Seeing  \\
       &  sec     &number& number  & arcsec  \\  
\hline
\hline
J      &   6      & 2    &   2     &  0.9    \\
H      &   4      & 1    &   2     &  0.8    \\
$K_S$  &   4      & 1    &   2     &  0.8    \\
\hline
\end{tabular}
{\small}
\end{table}

\subsection{Source Catalog}

It is well known that galactic morphology is correlated with the local density of galaxies \citep{Dressler1980} . 
Therefore, in galaxy clusters, there is also a correlation between morphology and the projected distance from 
the cluster center. In this sense \cite{WhitmoreGilmore1991} found that most of the variation occurs within
0.5 Mpc of the cluster center and that the effect is strongest in clusters with dominant galaxies, 
with the percentage of ellipticals rising from 18\% at 600 kpc from the cluster center to 65\% at distances of 100 kpc. 
Accordantly, \cite{Teu2003} fount that only within the central 200 kpc the E+S0 component of galaxy clusters become 
the dominant component.

Related with the identification of the red sequence, in order to reduce the relative contamination of field galaxies, 
\cite{Gla98} used only the inner 0.5$\mpc$ projected radius of the cluster centers to derived the slope of the 
red sequence for a sample of 50 galaxy clusters, spanning a redshift range $0<z<0.75$. Likewise, \cite{Stott09} consider 
galaxies within 0.6 Mpc radius of the cluster center limiting the contamination from field galaxies for the study of the 
evolution of the red sequence slope.

In order to minimize the contamination of field sources and to identified red-sequence elliptical galaxies, 
in this work we concentrated on the very central region around the X-ray surface brightness peak of Suzaku J1759$-$3450, 
which falls on VVV tile d261. We restricted our analysis to the inner $350 kpc$ of the cluster center that correspond 
to $~$1.5 core radius according to the $\beta$ model adopted for \cite{Mori13} to fit the X-ray profile of the 
surface brightness in the $0.5-10$keV band.

Although CASU generates photometric catalogues for all VVV data, we preferred to create a
new one, using SExtractor \citep{Bertin96}, because it is better 
suited to treat extended sources. We run SExtractor in double-image 
mode, with the K$_s$ deep image as a reference because of its better quality
and lower susceptibility to extinction. An object was considered 
detected if the flux exceeded at least twice the local background 
noise level, and spanned over at least ten connected pixels on the 
VVV images. The main entries in our final photometric catalog are:

{\it i)} $JHK_S$ magnitudes in three pixels radius apertures. This is recommended to 
represent the flux for all the objects in the field of the VIRCAM and to calculate colors. 
The aperture correction was applied to these magnitudes considering the 
{\it apcor}\footnote{http://casu.ast.cam.ac.uk/surveys-projects/vista/technical/catalogue-generation} 
parameter available in the header of the fits images. This correction, useful for point sources, 
also work well as a first order seeing correction for faint galaxies.

{\it ii)} $JHK_S$ total magnitudes. For these we used the MAG$\_$AUTO from SExtractor 
which is based by Kron's algorithm \citep{Kron80}.

{\it iii)} star-galaxy separation parameters:  CLASS$\_$STAR and half-light radius parameter, $r_{1/2}$.

2MASS-based CASU-estimated zero points were used to flux calibrate the images.
For the extinction correction we used the high resolution extinction map of the 
Milky Way from \citet{Gonza12}. This reddening map has a resolution of 2 arcmin and it 
is available via a web based tool BEAM (Bulge Extinction and Metallicity) 
calculator\footnote{http://mill.astro.puc.cl/BEAM/calculator.php}. This tool provides a lower limit
for the mean extinction, $A_{K_S}$ and the E(J-$K_S$) for a given set of coordinates and radius 
of the region. Finally, we adopted the \citet{Nishi09} extinction law obtaining the mean 
values of $A_J = 0.26$, $A_H=0.14$ and $A_{K_S}=0.09$.
For this analysis we used only objects with $K_S$=11-15.5\,mag to avoid 
saturated or too faint stars and blended objects.

Our final catalog contains 1308 sources in the central region of 
the X-ray emission. 
The photometric calibration was cross-checked comparing the 
magnitudes of the point sources with those of the matching 
CASU sources. The average precision of our magnitudes respect to the CASU ones is 
$\sim 0.01 mag$ in the three bands.

\section{Results}

\subsection{Star-Galaxy Separation}

\begin{figure}
\includegraphics[height=78mm,width=89mm]{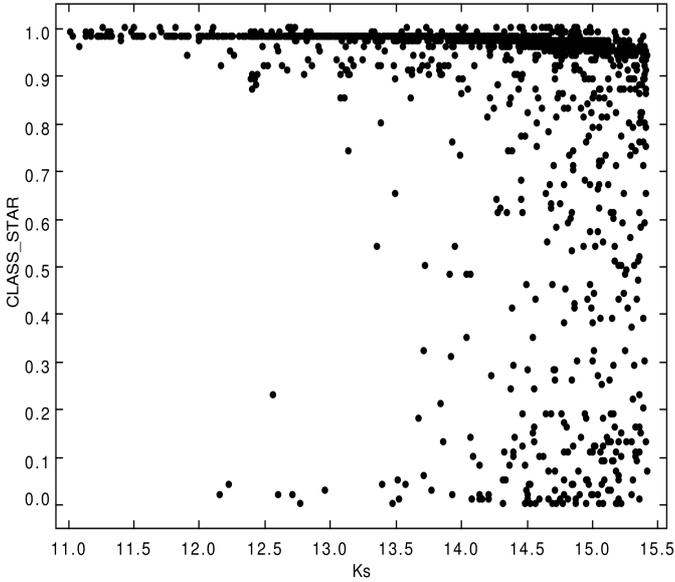}
\caption{The star-galaxy separation: the SExtractor stellarity 
index CLASS$\_$STAR versus the total $K_S$ band source magnitude.}
\label{fig:stelarity}
\end{figure}

Our star-galaxy classification was based on the SExtractor's 
stellarity index CLASS$\_$STAR since the
stars lie in a sequence near CLASS$\_$STAR$=1$, and galaxies 
near CLASS$\_$STAR$=0$. The two sequences merge for fainter 
sources, and \cite{Bertin96} showed that the stellarity index is 
strongly seeing dependent in this regime. 
In Fig.\,\ref{fig:stelarity} we show the relation between the CLASS$\_$STAR
parameter and the total $K_S$ magnitude for our dataset. Taking into 
account the presence of the two sequences, we removed star-like sources from our catalog 
adopting a conservative limit of CLASS$\_$STAR$<$0.5.

\begin{figure*}
\includegraphics[width=190mm,height=85mm]{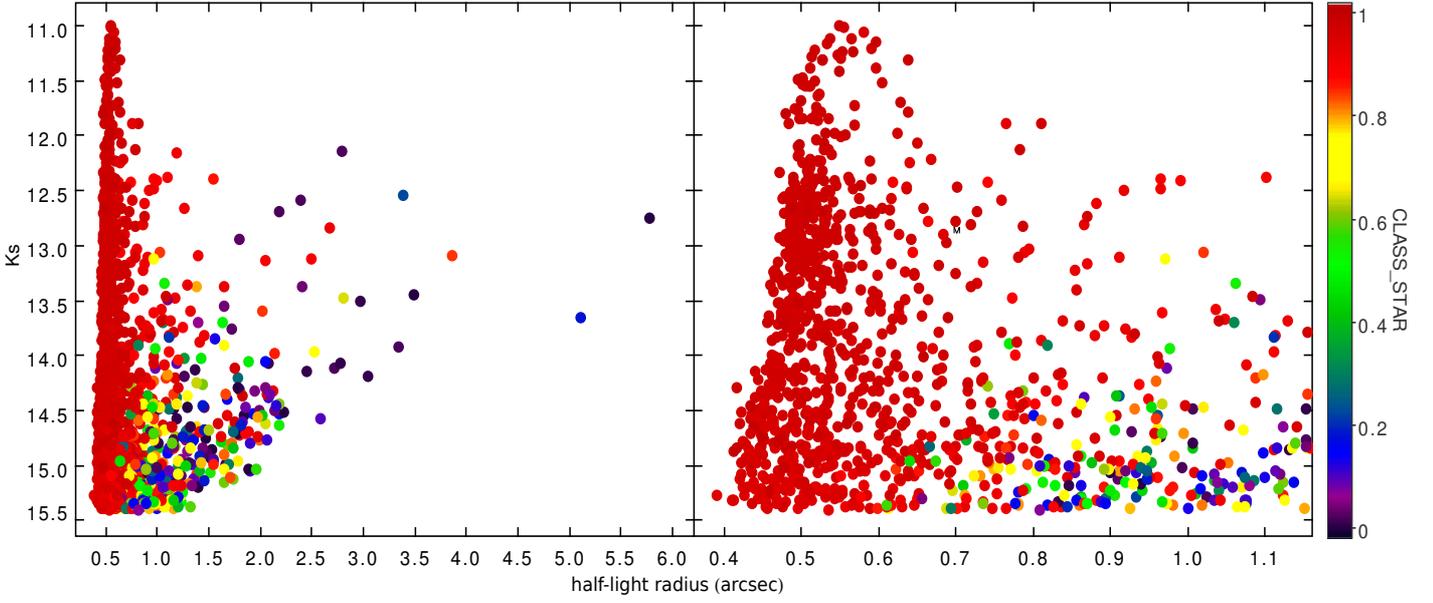}
\caption{Left: Half-light size $r_{1/2}$ versus $K_S$ magnitude diagram for the whole sample of detected 
sources in the central region. 
Right: Enlargement of the left panel for low values of $r_{1/2}$ radius. The points are color-codded 
according to the SExtractor's stellarity parameter CLASS$\_$STAR (the bar on the right).
}
\label{fig:svmd}
\end{figure*}

The size versus magnitude diagram is another useful tool in order to perform 
star-galaxy separation. For this purpose we used the half-light radius parameter, $r_{1/2}$.
that measures the radius that encloses 50 \% of the object’s 
total flux. For objects larger than the seeing, $r_{1/2}$ 
is independent of magnitude, placing the extended objects in a 
well-separated region of the diagram. 
The left panel of Fig.\,\ref{fig:svmd} shows the half-light radius versus total $K_S$ 
magnitude for the whole sample of sources detected in the central region around the X-ray peak emission. 
The points are color-codded according to the stellarity
parameter CLASS$\_$STAR (the bar on the right) to underline the
consistency of the two methods: the majority of the objects with 
CLASS$\_$STAR$\sim$1 approximately populate the stellar locus at 
$r_{1/2}$$<$0.7\,arcsec (which corresponds to $\sim$ 2 pixel radius) while that sources with 
CLASS$\_$STAR$<$0.5 tend to have larger $r_{1/2}$ as can be observed in the right panel of 
Fig.\,\ref{fig:svmd}.

Combining the two methods, we define as galaxy candidates those 
objects with CLASS$\_$STAR$<$0.5 and $r_{1/2}>$0.7\,arcsec, obtaining 
a sample of 187 galaxy-like sources. Many of these objects could be unresolved double 
or multiple stars, because of the extreme crowding in the Milky
Way bulge. So, in the next section, colors are used to improve the galaxy candidate 
selection method mainly for those galaxies belonging to galaxy clusters.

\subsection{Red-Sequence Galaxies}

Galaxy clusters contain well-defined, highly regular population of 
elliptical and lenticular galaxies, that form a structure in their 
color-magnitude diagrams, known as red-sequence \citep{GY00}.
This alternative 
approach to identify cluster of galaxies has been proven to be highly successful by 
numerous authors  \citep{GY00,LC04,ilona06} and we  used it for a suitable
identification of cluster members.
 
\begin{figure}
\includegraphics[width=95mm,height=85mm]{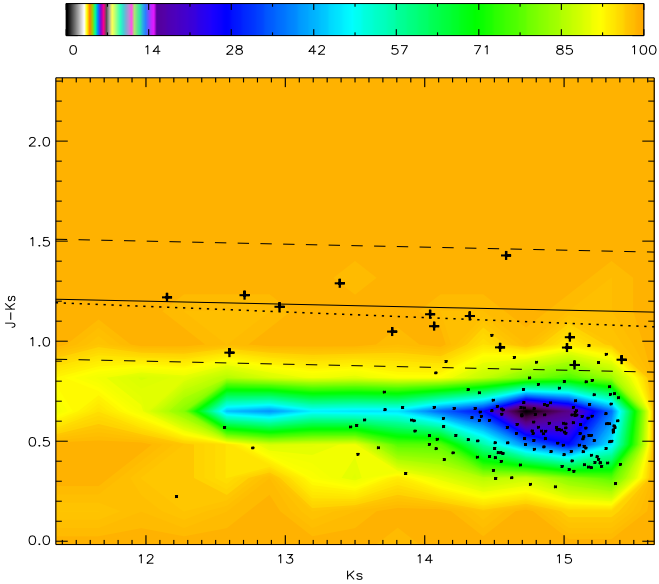}
\caption{Color-magnitude diagram $J-K_S$ versus $K_S$. 
The density map shows the whole sample in the central region of the cluster 
(color scale corresponding to the percentage of objects, respect to the total number, as shown in the key).
The red-sequence model is represented by a solid line and the $\pm 5\sigma$ around the model by dashed lines, 
respectively. The dots correspond to the sources that met the selection criteria of Section 3.1. 
The crosses mark the red-sequence galaxies with visual confirmation. The red sequence, fitted to the bright 
galaxy cluster member candidates, is shown by the dotted line. 
}
\label{fig:cmd}
\end{figure}

To further strengthen the identification of the galaxy cluster member candidates 
we used a color selection criteria based on a red-sequence model derived from 
\cite{BC03}. The adopted model uses the Padova 94 \citep{Pad94a,Pad94b} evolutionary 
tracks and the \cite{Chab03} initial mass function with a passively evolving, instantaneus-burst 
stellar population with a redshift formation of $z=3$.
The obtained mean red-sequence model is shown in Fig.\,\ref{fig:cmd} (black solid line) where the 
$J-K_S$ versus $K_S$ color-magnitude diagram of our field is shown. It 
was computed averaging the models for different metallicities at the redshift of the galaxy cluster, 
$z=0.13$.
  
Based on the aforementioned, we defined as cluster member those galaxies within $\pm 5\sigma$ 
(dashed lines in Fig.\,\ref{fig:cmd}) from the adopted red-sequence model in the color-magnitude diagram. 
It is important to emphasize that color gradients could vary from one
galaxy to another, so to obtain unbiased colors of the galaxy candidates
we used fixed aperture magnitudes for all filters.
This color cut yielded 22 final galaxy candidates.

The density map in the color-magnitude diagram of Fig.\,\ref{fig:cmd} represents the whole sample of 
objects in the 
central region of the X-ray peak emission. On the other hand the points in the figure indicate those 
source satisfying the CLASS$\_$STAR$<$0.5 and $r_{1/2}>$0.7\,arcsec criteria. From this figure is 
possible to observe two well defined region in the color-magnitude diagram: 
one horizontal sequence at $J-K_S<$1.0 (shown by the color density map) corresponding to the stellar 
sequence of galactic stars, probably with types later than G5 and earlier than K5 \citep{Fin00}, 
 and the other one corresponding to the objects defined as galaxy candidates  
located mainly in the region close to the modeled red sequence (cross symbols).
The candidate galaxies in the region of the red-sequence could have an appearance corresponding to
early type objects such as elliptical and lenticular galaxies.

Next, we inspected visually these galaxy candidates on a false-color 
multiband image (built from J, H, Ks frames) using Aladin \citep{bon00}, 
to identify extended sources with the typical galactic features:
color, sizes and ellipticity.
We verified the galactic features of 15 candidates whereas 7 remained unclassified,
typically because they were too faint and an objective classification is not possible.
Moreover, from these 15 objects we classified them in two categories: Type I sources as those 
with clear galaxy features resulting 12 candidate galaxies in this category; and Type II the 
remaining sources with not completely clear features expected for galaxies.
The visually confirmed galaxies belonging to the red-sequence can be seen in
Fig.\,\ref{fig:cmd} (cross symbols).

\begin{figure}
\includegraphics[width=95mm,height=85mm,angle=0]{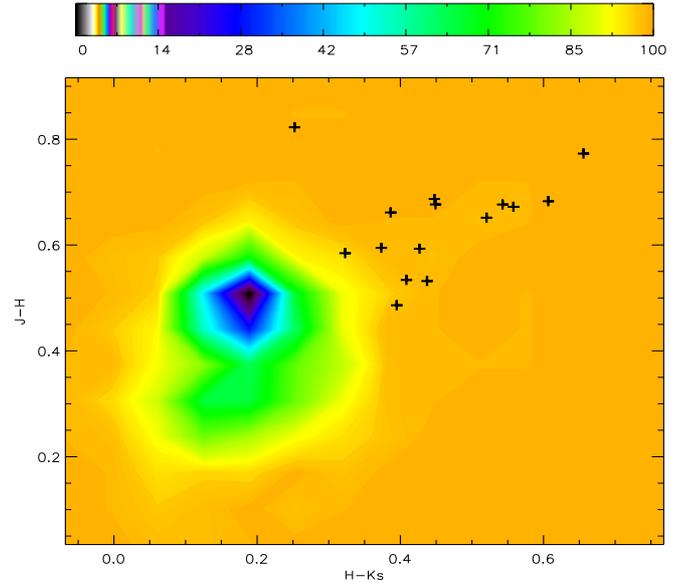}
\caption{Color-color diagram $J-H$ versus $H-K_S$. 
The density map shows the whole sample in the central region of the cluster 
(color scale corresponding to the percentage of objects, respect to the total number, as shown in the key).
 The crosses mark the red-sequence galaxies, selected from Fig.\,\ref{fig:cmd}.
}
\label{fic:ccd}
\end{figure}

The cluster member galaxy candidates are, indeed, red in all 
colors, as can be seen from the color-color diagram shown in 
Fig.\,\ref{fic:ccd} (crosses) clearly, they occupy a different locus than the 
foreground stars. This is consistent with the 
results of \cite{Amores12} who visually identify galaxies behind 
the Galactic plane in the VVV survey, and used color-color 
diagrams to confirm the galaxy-star separation.

\begin{figure*}
\includegraphics[width=155mm,height=230mm ]{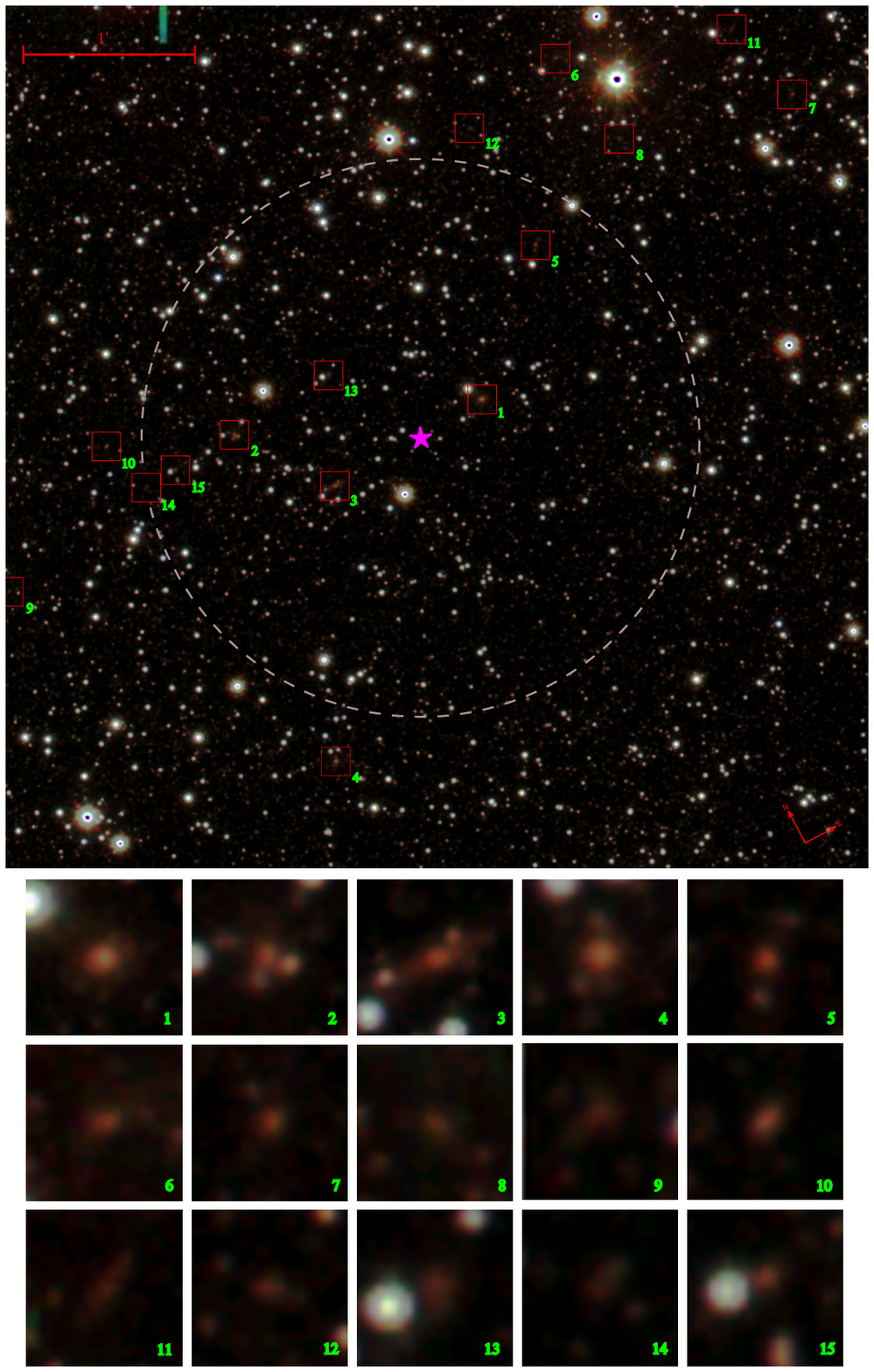}
\caption{False-color $J$ (blue), $H$ (green) and $K_S$ (red) 
image of the cluster region. 
The star symbol indicates the peak 
of the X-ray emission. All the candidate galaxies are enumerated in the images.
The grey dashed line circle shows the area corresponding to the core radius of the X-ray emission, 
$r_c=1.61 arcmin$. A scale bar is shown at the left top.
The postage stamp images of every galaxy candidate, with the identification number, are at the bottom.
}
\label{fig:map}
\end{figure*}

Finally, the Fig.\,\ref{fig:map} shows a false-color multiband (J, H, Ks) image, centered in the X-ray peak emission, 
where the five brightest galaxies can be found very close to this peak.
From this image, it is possible to observe that galaxy cluster member candidates show redder 
colors, with respect to the 
foreground stars, and extended morphology characteristic of galaxies.
Besides, the area corresponding to core radius according to the $\beta$ model adopted for \cite{Mori13} 
to fit the X-ray profile of the surface brightness is shown.

Table\,2 lists the cluster galaxy candidate parameters ordered by $Ks$ total magnitude.
The list includes an identification number, 
coordinates, aperture and total magnitudes and classification.

Although we can not estimate the real contamination of background galaxies without redshift measurements, 
the fact that the restriction of the analysis to the central region of the galaxy cluster favors elliptical 
and S0 galaxies with typical red colors, minimizes the background galaxy contamination. 
For instance, the red sequence has been used as an efficient method to detect galaxy clusters, at optical 
and infrared wavelengths,  with a contamination of less than 5\% \citep{GY00}. More recently, 
\citet{Barkhouse06} uses the red sequence to improve the Voronoi tessellation and percolation technique 
reducing significantly the contamination.

\begin{table*}
\center
\caption{Cluster Galaxy Candidates. J($3$), H($3$), $K_S$($3$) are magnitudes in apertures of 3 pixels of 
radius and $JHK_S$ correspond to total magnitudes. The class parameter indicates clear (I) or unclear (II) 
galaxy features.}
{\small}
\begin{tabular}{|c c c c c c c c c c| }
\hline
ID & RA(J2000)& DEC(J2000)& J($3$)  &  H($3$) &  $K_S$($3$) &  J &  H &  $K_S$  & class \\
   & hh:mm:ss.ss & dd:mm:ss.ss & mag  &  mag &  mag &  mag &  mag &  mag  &  \\
\hline
\hline
 1 &  17:59:19.44 &  -34:50:18.7 &  15.19 $\pm$  0.03  &  14.52 $\pm$ 0.02  &  13.98 $\pm$ 0.02  &  13.00 $\pm$ 0.02  &  12.50 $\pm$ 0.02  &  12.12 $\pm$ 0.02 &   I  \\ 
 2 &  17:59:12.98 &  -34:49:48.5 &  15.30 $\pm$  0.03  &  14.76 $\pm$ 0.02  &  14.36 $\pm$ 0.02  &  13.39 $\pm$ 0.02  &  12.95 $\pm$ 0.02  &  12.62 $\pm$ 0.02 &   I  \\ 
 3 &  17:59:14.75 &  -34:50:19.9 &  15.64 $\pm$  0.04  &  14.97 $\pm$ 0.03  &  14.41 $\pm$ 0.03  &  13.40 $\pm$ 0.02  &  13.06 $\pm$ 0.02  &  12.71 $\pm$ 0.02 &   I  \\   
 4 &  17:59:10.97 &  -34:51:42.2 &  15.50 $\pm$  0.03  &  14.85 $\pm$ 0.02  &  14.33 $\pm$ 0.02  &  13.70 $\pm$ 0.02  &  13.27 $\pm$ 0.02  &  12.96 $\pm$ 0.02 &   I  \\ 
 5 &  17:59:22.85 &  -34:49:42.0 &  16.21 $\pm$  0.06  &  15.53 $\pm$ 0.04  &  14.92 $\pm$ 0.04  &  14.12 $\pm$ 0.03  &  13.77 $\pm$ 0.03  &  13.39 $\pm$ 0.03 &   I  \\ 
 6 &  17:59:25.91 &  -34:48:49.5 &  16.21 $\pm$  0.05  &  15.55 $\pm$ 0.04  &  15.16 $\pm$ 0.04  &  14.91 $\pm$ 0.06  &  14.30 $\pm$ 0.05  &  13.77 $\pm$ 0.04 &   I  \\ 
 7 &  17:59:31.11 &  -34:49:39.5 &  16.28 $\pm$  0.06  &  15.60 $\pm$ 0.04  &  15.15 $\pm$ 0.04  &  14.58 $\pm$ 0.04  &  14.22 $\pm$ 0.04  &  14.04 $\pm$ 0.05 &   I  \\ 
 8 &  17:59:26.33 &  -34:49:24.0 &  16.53 $\pm$  0.07  &  15.70 $\pm$ 0.05  &  15.45 $\pm$ 0.05  &  14.69 $\pm$ 0.04  &  13.99 $\pm$ 0.03  &  14.07 $\pm$ 0.05 &   I  \\   
 9 &  17:59:05.36 &  -34:49:56.9 &  16.73 $\pm$  0.08  &  16.05 $\pm$ 0.06  &  15.60 $\pm$ 0.06  &  15.24 $\pm$ 0.06  &  14.63 $\pm$ 0.05  &  14.32 $\pm$ 0.05 &  II  \\ 
10 &  17:59:09.71 &  -34:49:30.0 &  15.94 $\pm$  0.05  &  15.40 $\pm$ 0.04  &  14.97 $\pm$ 0.04  &  15.35 $\pm$ 0.05  &  14.82 $\pm$ 0.05  &  14.54 $\pm$ 0.05 &   I  \\
11 &  17:59:30.54 &  -34:49:10.0 &  17.01 $\pm$  0.10  &  16.24 $\pm$ 0.07  &  15.58 $\pm$ 0.06  &  15.58 $\pm$ 0.07  &  15.14 $\pm$ 0.07  &  14.58 $\pm$ 0.06 &   I  \\ 
12 &  17:59:22.85 &  -34:48:56.2 &  16.47 $\pm$  0.06  &  15.88 $\pm$ 0.05  &  15.51 $\pm$ 0.05  &  15.62 $\pm$ 0.06  &  15.22 $\pm$ 0.06  &  15.02 $\pm$ 0.07 &  II  \\ 
13 &  17:59:16.07 &  -34:49:45.6 &  16.18 $\pm$  0.05  &  15.59 $\pm$ 0.04  &  15.16 $\pm$ 0.04  &  16.05 $\pm$ 0.06  &  15.45 $\pm$ 0.05  &  15.04 $\pm$ 0.05 &   I  \\
14 &  17:59:10.11 &  -34:49:49.2 &  16.53 $\pm$  0.07  &  16.05 $\pm$ 0.06  &  15.65 $\pm$ 0.06  &  15.53 $\pm$ 0.07  &  15.28 $\pm$ 0.08  &  15.07 $\pm$ 0.09 &  II  \\   
15 &  17:59:11.06 &  -34:49:48.5 &  16.55 $\pm$  0.07  &  15.97 $\pm$ 0.06  &  15.65 $\pm$ 0.06  &  16.42 $\pm$ 0.07  &  16.01 $\pm$ 0.06  &  15.41 $\pm$ 0.06 &   I  \\ 

\hline
\end{tabular}
\end{table*}

\section{Discussion}

The analysis reported in the previous sections demonstrated that 
the field of Suzaku J1759$-$3450 contains multiple galaxy 
candidates that show the colors and magnitudes cluster members,
including the brightest galaxy, with $K_S \sim$ 12.1$\,mag$, and 
four other prominent objects in the magnitude range 12.5 $<K_S<$ 13.5. 
\citet{Brough02} indicates that the brightest cluster galaxies 
(BCGs) in clusters with X-ray luminosity
$L_x$(0.3-3.5\,keV)$>$1.9$\times$10$^{44}$erg s$^{-1}$ have 
apparent magnitudes around $K_S$$\sim$12\,mag, consistent with 
our measurements.
Moreover, the apparent magnitude of the BCG is consistent with an absolute 
magnitude of $M_{K_S}$ $\sim -$ 25.6, if we assume that this 
galaxy is  a member of a galaxy cluster at redshift $z$=0.13, 
also in agreement with \cite{Brough02}. These results reinforce 
the assumption that the red-sequence detected indeed 
corresponds to the new galaxy cluster, reported by \cite{Mori13}.

Considering the visual appearance of the BCG corresponding to a very 
luminous elliptical galaxy,  
half magnitude brighter than the second brightest galaxy, the large diffuse halo, and 
the proximity to the X-ray emission peak  ($\sim$25\,arcsec) we could speculate that 
this is a cD galaxy.

In addition to the modeled red-sequence, we also calculated the red-sequence slope using only the 
visually 
confirmed galaxies brighter than $K_S=$14.1, corresponding to galaxies 2 magnitudes fainter than the 
brightest one to minimizing the contamination which is more likely for the faintest galaxies.
This empiric red-sequence is shown in Fig. 3 (dotted line) showing a good agreement with the modeled one.
It is important to highlight that a random selection of galaxies at
different redshifts will not form a coherent red sequence \citep{GY00}, therefore
the projection effect would not affect significantly this selection criteria.
Anyway spectroscopic measurement 
of the redshifts of the newly identified cluster members will be extremely useful to confirm these results.
The obtained value of the slope, $k_{JK_S} = -0.028 \pm 0.011$ is consistent with that found by \cite{Stott09} for 
three galaxy clusters at $z\sim$0.13.

It is worth mentioning that the detected candidates to galaxy cluster member have an 
asymmetric spatial distribution with respect to the position of the X-ray peak emission.
Instead, the galaxies are preferentially located on one side of the X-ray peak emission.
In this sense, \cite{FJ82} and \cite{JF84}
propose a two-dimensional classification scheme for the X-ray morphology of 
the intracluster gas which represents a sequence of cluster evolution. Moreover, the presence or absence of a central, dominant galaxy in the cluster is considered in 
the classification scheme of \cite{FJ82}. 
The authors classify the clusters as being early if the overall X-ray surface brightness 
distribution is irregular or evolved if it is regular. 
In addition, clusters containing a central dominant galaxies are classified as X-ray dominant (XD) while those without such a galaxy are classified as non-X-ray dominant (nXD). The nXD clusters have 
larger X-ray core radii and there is no strong X-ray emission associated with any individual galaxy 
in these systems. 
That is, the galaxy cluster could be X-ray dominant or non-X-ray dominant for the regular or irregular X-ray surface brightness.

As it was mentioned, in this particular case, the cluster has a regular X-ray morphology without a central dominant galaxy in the X-ray peak emission. So, Suzaku J1759$-$3450 can be classified as a regular nXD cluster.
Several galaxy clusters have a similar characteristic i.e. A1656 \citep{Abramopoulos81}, A576 \citep{WS80},
A2255, A2256 \citep{FJ82} and CA0340-538 \citep{Ku83}. This feature indicates that the galaxy cluster is still
evolving and suffering dynamical processes. 
In the same sense \cite{Mori13} found that, even with a approximately circular X-ray morphology, 
the peak position of the surface brightness is slightly shifted from the center of the circular emission 
in the south-east direction suggesting that the cluster dynamical relaxation has not taken place yet.

\section{Conclusions}

Summarizing, we used VVV Survey data to obtain deep NIR photometry of objects 
in the vicinity of the recently identified galaxy cluster Suzaku 
J1759$-$3450 at $z$=0.13. At least fifteen galaxies were detected 
within a projected distance of 350$kpc$ from the central peak 
of the X-ray peak emission, and five of them are extremely bright.
 
All these cluster member candidates have the typical colors and magnitudes of galaxies at 
redshift $z$=0.13. The candidate BCG is more than half magnitude 
brighter than the next brightest galaxy candidate, indicating 
that it may be a cD galaxy. These results lead us to conclude 
that Suzaku J1759$-$3450 is indeed an obscured galaxy cluster at 
$z$=0.13. Spectroscopic observations are needed to measure the 
redshifts of the newly identified cluster members, and to study 
their stellar populations in detail.

\begin{acknowledgements}
We would like to thanks to anonymous referee for the comments that helped to improve the paper.
GC, SA and FD acknowledge the support from CONICET through grant PIP 2012-2014,
GI 11220110100298. 
Support for MH has been provided by the BASAL Center for Astrophysics and
Associated Technologies PFB-06, the FONDAP Center for Astrophysics N. 15010003,
and the Ministry for the Economy, Development and Tourism’s Programa Iniciativa
Cient\'ifica Milenio through grant P07-021-F, awarded to The Milky Way Millenium Nucleus.
DM is supported by  Project IC120009 Millennium Institute of Astrophysics (MAS) of Iniciativa Cient\'ifica Milenio del Ministerio de Econom\'ia, Fomento y Turismo, by the Basal Center for Astrophysics and Associated Technologies (CATA) PFB-06, and by Fondecyt Project No. 1130196.
GV acknowledge the support from Agencia Nacional de Promoci\'on Cient\'ifica y Tecnol\'ogica, PICT 2010 Bicentenario N°0680.

\end{acknowledgements}

\end{document}